# Spin relaxation of localized electrons in n-type semiconductors

**K.V.Kavokin**

*A.F.Ioffe Physico-Technical Institute, Polytechnicheskaya 26, 194021 St-Petersburg, Russia*
e-mail: kidd.orient@mail.ioffe.ru

Abstract: The mechanisms that determine spin relaxation times of localized electrons in impurity bands of n-type semiconductors are considered theoretically and compared with available experimental data. The relaxation time of the non-equilibrium angular momentum is shown to be limited either by hyperfine interaction, or by spin-orbit interaction in course of exchange-induced spin diffusion. The energy relaxation time in the spin system is governed by phonon-assisted hops within pairs of donors with an optimal distance of about 4 Bohr radii. The spin correlation time of the donor-bound electron is determined either by exchange interaction with other localized electrons, or by spin-flip scattering of free conduction-band electrons. A possibility of optical cooling of the spin system of localized electrons is discussed.

1) *Introduction.*

Strange though it may seem, 40 years of research (since the pioneering work by G.Lampel [1]) on optical orientation of electron and nuclear spins have not filled all the major blank spaces in this very interesting area of physics. Answers to many questions of primary importance are being approached just now. One of these actively developed fields with a long history is the problem of spin memory – not only in nanostructures brought forth by sophisticated novel technologies, but also in bulk semiconductor crystals. A part of this broad field, the physics of relaxation processes in the spin system of interacting localized electrons in non-magnetic n-type semiconductors, is the subject of this paper. It consists of an extended theoretical introduction – partly based on published results, partly original – followed by a survey of relevant experiments. The consideration is limited to direct-gap semiconductors, like GaAs, which are mainly studied in the experiments on optical orientation. Aiming at bringing together experiment and theory, I will concentrate at bulk crystals where localizing potentials and the concentration of localization centres are determined by doping and can be easily calculated. The concentrations and temperatures considered are those low enough for electrons to remain localized.

2) *Relaxation time scales in spin systems.*

Non-equilibrium spin polarization created by an external pumping (for instance, by circularly polarized light) persists during some characteristic time after switching off the pump. This time is often called the spin memory time. It should be, however, noted that relaxation processes in spin systems are not, generally, characterized by a single time scale. Depending on the experimental conditions, the observed "spin memory time" can be determined by different relaxation processes.

These may include, for instance, relaxation of the components of the vector of total angular momentum (the spin relaxation proper) or of the energy of the spin system. The relaxation of a component of the angular momentum along a certain axis requires only that the spin interactions lack symmetry with respect to rotation about this axis. The energy relaxation of the spin system also requires coupling of this system with an energy reservoir of sufficient capacity (phonons, for instance). Clearly, the relaxation of angular momentum and the relaxation of energy can be provided by quite different interactions, resulting in disparate scales of corresponding relaxation times. For example, the relaxation of the angular momentum of nuclear spins in a solid is provided by their magneto-dipole interaction and occurs during $10^{-4}$ c, while the energy transfer from the nuclear spin system to the lattice may take hours [2, 3]. The difference of relaxation times for angular momentum and energy is usually unsubstantial for free electrons, because in this case main spin relaxation mechanisms are powered by the motion of electrons, very efficiently transforming the energy of spin interactions into the kinetic energy. As we shall see, for localized electrons this difference may be of primary importance.

There is also a difference between relaxation times of longitudinal and transverse spin components in a magnetic field. The latter is important for interpretation of experiments on "spin beats" and magnetic resonance, which are outside the scope of this paper. The longitudinal time is normally longer than the transversal time for two reasons. One is that the relaxation of the longitudinal spin component is accompanied by dissipation of the Zeeman energy, which requires coupling of the spin system with the lattice. The second reason for this difference is that the Zeeman splitting slows down transitions between spin sublevels of individual spins.

In the following, we will specify which sort of relaxation is discussed each time when such a difference may occur.

3) *Specifics of spin relaxation in n-type semiconductors.*

In n-type semiconductors, the electron spin relaxation is usually distinctively slower than in p-type ones. The main reason is the absence of the powerful relaxation channel due to exchange scattering of electrons by holes (Bir-Aronov-Pikus mechanism). At high temperatures, the spin relaxation is dominated by the Dyakonov-Perel mechanism that involves the electron spin precession in effective magnetic fields of a spin-orbit nature, arising when the electron moves. Such spin-orbit fields exist in semiconductors and semiconductor structures lacking the inversion symmetry. They are proportional to odd powers of the electron wave vector components (third power in bulk cubic crystals and, typically, first in two-dimensional structures). Their direction is determined by directions of the wave vector and of the crystal axes. For example, the spin-orbit field in zinc-blend crystals is given by the expression

$$B_x^{SO} = \alpha_{SO} \hbar^3 \left( \mu_B g m \sqrt{2mE_g} \right)^{-1} k_x \left( k_y^2 - k_z^2 \right) \qquad (1)$$

(other components are obtained by permutation of indices). Here $m$ is the electron effective mass, $E_g$ is the band gap, $k$ is the electron wave vector, $\mu_B$ is the Bohr magneton, $g$ is the conduction-band electron g-factor, and $\alpha_{SO}$ is a dimensionless constant. For GaAs, $\alpha_{SO} \approx 0.07$ [4, 5].

When the electron is scattered, the direction of the spin-orbit field is changed randomly. If, while the electron spin turns around the spin-orbit field, the field changes its direction many times, the spin relaxation time $\tau_S$ is given by the following formula:

$$\frac{1}{\tau_S} = \langle \Omega^2 \rangle \tau_c \qquad (2)$$

where $\langle \Omega^2 \rangle = \langle (\mu_B g B^{SO}/\hbar)^2 \rangle$ is the mean squared Larmor frequency of the electron spin in the spin-orbit field, and $\tau_c$ is the correlation time of this field, proportional to the momentum relaxation time of electrons [6]. Equation (2) is valid also for Fermi-edge electrons in a degenerate semiconductor; in this case $\langle \Omega^2 \rangle$ is determined by the Fermi momentum [7, 8]. However, lowering the temperature results in a quite different behaviour in non-degenerate semiconductors. In that case, the concentration of electrons in the conduction band becomes very small, the electrons being bound to localization centres (in bulk crystals, to donor impurities). The ensemble of localized electron states is often called the impurity band. Bound states have zero average wave vectors, and for this reason the Dyakonov-Perel mechanism does not work for localized electrons. Still, as we shall see, the spin-orbit interaction remains a major cause of spin relaxation for impurity-band electrons.

4) *Spin-orbit interaction and the asymptotic form of the donor wave function.*

The spin-orbit interaction does not cause spin relaxation of a single localized electron, because, due to the Kramers theorem, it does not split the electron spin sublevels. Combined with the electron-phonon interaction, the spin-orbit coupling can induce spin-flip transitions, but at liquid-helium temperatures their intensity is fairly low; according to Khaetskii and Nazarov [9], in quantum dots the corresponding spin relaxation times exceed 0.1 s. At the same time, the structure of the electron wave function changes: it becomes a spinor whose components with different spin indices are different functions of coordinates. This fact is of primary importance for spin relaxation in the impurity band, and we will discuss it in detail. As mentioned above, in semiconductors lacking the inversion symmetry the spin-orbit interaction results in the appearance of spin-dependent terms in the Hamiltonian of the conduction band, having the general form $\mu_B g \vec{B}^{SO}(\vec{k}) \cdot \vec{S}$, where the effective spin-orbit field $\vec{B}^{SO}(\vec{k})$ is an odd function of the wave vector components. Because of their smallness, these terms have practically no effect on the binding energy and the wave function shape near the localization centre. But the behaviour of the wave function at large distances is seriously changed. The asymptotic form of the wave function far away from the centre, where the localizing potential is close to zero, can be obtained in the quasi-classical approximation [10]. The wave function at a distance larger than some (arbitrary) $r_0$ can be approximately written as

$$\Psi \propto \exp\left(\frac{iS(\vec{r})}{\hbar}\right) \qquad (3)$$

where $S(\vec{r}) = \hbar \int_{r_0}^{r} \vec{k} d\vec{r}'$ is the action, the integral is along the straight trajectory emerging from the centre, and the wave vector should be found using the condition

$$\frac{(\hbar k)^2}{2m} + \mu_B g \vec{B}^{SO}(\vec{k}) \cdot \vec{S} = E - U(\vec{r}) \tag{4}$$

At large distances from the centre the potential energy $U(\vec{r})$ can be neglected. Using the smallness of the spin-orbit terms, we can seek $k$ in the form: $k = k_0 + \Delta k$, where $k_0 = \sqrt{\frac{2m(E-U(r))}{\hbar^2}} \approx i\sqrt{\frac{2mE_B}{\hbar^2}}$, $E_B$ is the electron binding energy. The spin-orbit correction $\Delta k$ is found from the condition $\frac{\hbar^2 k_0 \Delta k}{m} + \mu_B g \vec{B}^{SO}(\vec{k}_0) \cdot \vec{S} = 0$, yielding $\Delta k = \frac{m \mu_B g \vec{B}^{SO}(\vec{k}_0) \cdot \vec{S}}{\hbar^2 k_0}$. As $\vec{B}_{SO}$ is an odd function of the wave vector, $\Delta k$ is a real number. Finally we obtain

$$\Psi \propto \exp\left(-\sqrt{\frac{2mE_B}{\hbar^2}} r\right) \exp\left(i \frac{m \mu_B g \vec{B}^{SO}\left(\sqrt{\frac{2mE_B}{\hbar^2}} \frac{\vec{r}}{r}\right) \cdot \vec{S}}{\hbar \sqrt{2mE_B}} r\right) \tag{5}$$

The spin-dependent multiplier in this expression has a form of the operator of finite rotation for the spin *S*. For this reason, if near the centre the localized electron state corresponds to a certain spin direction (i.e. its spin projection on a certain axis equals ½), then at the distance r from the centre the spin will have the same projection on the axis turned through the angle [11]:

$$\gamma(r) = \frac{\hbar m \mu_B g \vec{B}^{SO}\left(\sqrt{\frac{2mE_B}{\hbar^2}} \frac{\vec{r}}{r}\right) r}{\sqrt{2mE_B}} \tag{6}$$

around the spin-orbit field $\vec{B}_{SO}\left(\sqrt{\frac{2mE_B}{\hbar^2}} \frac{\vec{r}}{r}\right)$. One can define the *spin-orbit length* $L_{SO}$ by the condition $\sqrt{\langle [\gamma(L_{SO}\vec{\lambda})]^2 \rangle} = 1$, where the angle is averaged over the directions of a unit vector $\vec{\lambda}$. For GaAs, the spin-orbit length is about 5μm. Using this parameter, one can write the angle-averaged rotation angle as

$$\langle \gamma^2(r) \rangle^{1/2} = r / L_{SO} \tag{7}$$

6) ***Spin rotation of the electron hopping from donor to donor, and anisotropic exchange interaction.***
    The asymptotic behaviour of the wave function of the localized electron affects the spin dynamics if there is more than one localization centre for the electron. Consider two centres (*i*) and

(*j*), situated not very far from each other. The wave functions of electrons localized at these two centres are $\Psi_i^\alpha = F(|\vec{r} - \vec{R}_i|)\exp(i\vec{\gamma}(\vec{r} - \vec{R}_i)\vec{\sigma}_{\alpha\beta}/2)\chi_\beta$ and $\Psi_j^\alpha = F(|\vec{r} - \vec{R}_j|)\exp(i\vec{\gamma}(\vec{r} - \vec{R}_j)\vec{\sigma}_{\alpha\beta}/2)\chi_\beta$, where $\vec{R}_i$ and $\vec{R}_j$ are position vectors of the two ions, $\vec{\gamma}(\vec{r}) = \gamma(\vec{r})\vec{B}_{SO}(\vec{r})/|\vec{B}_{SO}(\vec{r})|$, $\vec{\sigma}_{\alpha\beta} = (\sigma_{\alpha\beta}^x, \sigma_{\alpha\beta}^y, \sigma_{\alpha\beta}^z)$ is a vector of the Pauli matrices, and $\chi$ is an eigenfunction of spin 1/2. One can see that $\Psi_i^{1/2}$ and $\Psi_j^{-1/2}$ are no longer orthogonal. Therefore, the electron tunnelling between these two centres may be accompanied by a spin flip.

The functions $\Psi_i^{1/2}$ and $\Psi_j^{-1/2}$ have one remarkable property. Let us choose different systems of spinor indices for centres (*i*) and (*j*), defining them by the relation $\chi_\alpha^j = \exp(i\vec{\gamma}(\vec{R}_{ij})\vec{\sigma}_{\alpha\beta})\chi_\beta^i$, where $\vec{R}_{ij} = \vec{R}_j - \vec{R}_i$. This simply means using different coordinate frames for spins at the two centres. These frames are transformed into each other by rotation through the angle $\gamma(\vec{R}_{ij})$. Under this choice, the two functions, $\Psi_i^{\alpha'} = F(|\vec{r} - \vec{R}_i|)\exp(i\vec{\gamma}(\vec{r} - \vec{R}_i)\vec{\sigma}_{\alpha'\beta}/2)\chi_\beta^i$ and $\Psi_j^\alpha = F(|\vec{r} - \vec{R}_j|)\exp(i\vec{\gamma}(\vec{r} - \vec{R}_j)\vec{\sigma}_{\alpha\beta}/2)\chi_\beta^j = F(|\vec{r} - \vec{R}_j|)\exp(i\vec{\gamma}(\vec{r} - \vec{R}_i)\vec{\sigma}_{\alpha\beta}/2)\chi_\beta^i$, will regain the spin orthogonality along the line connecting the two centres. All the overlap integrals entering the tunnelling matrix elements are governed by the narrow region along this line, where the product of the two wave functions is the largest. For this reason, the electron with the spin index +½ at one centre will, after tunnelling to the other centre, remain in the +1/2 spin state, but in a rotated frame. In other words, the main effect of tunnelling from (*i*) to (*j*) on the electron spin is just turning the spin through the angle $\gamma(\vec{R}_{ij})$.

If the centre (*j*) is occupied by another electron, the electrons will be coupled by the exchange interaction. Since the exchange integrals are also governed by the "tunnelling corridor" along the straight line connecting the centres, the effect of the spin-orbit interaction is very similar to that in the case of tunnelling: the exchange interaction will now couple spin operators defined in different coordinate frames: $\hat{H}_{ex} = 2J_{ij}\vec{S}'_i \cdot \vec{S}'_j$ [12]. As shown in [13], in the case of linear in *k* spin-orbit terms this result is exact for any distance between localization centres, not only in the asymptotic region. Transforming the spin operators back to the laboratory frame, we obtain the following expression for the exchange Hamiltonian [12]:

$$\hat{H}_{ex} = 2J_{ij}\vec{S}'_i \cdot \vec{S}'_j = 2J_{ij}\left[\cos\gamma_{ij}\vec{S}_i \cdot \vec{S}_j + \sin\gamma_{ij}\frac{\vec{\gamma}_{ij}}{\gamma_{ij}} \cdot \vec{S}_i \times \vec{S}_j + (1 - \cos\gamma_{ij})\left(\frac{\vec{\gamma}_{ij}}{\gamma_{ij}} \cdot \vec{S}_i\right)\left(\frac{\vec{\gamma}_{ij}}{\gamma_{ij}} \cdot \vec{S}_j\right)\right] \quad (8)$$

Here the first term is the scalar exchange interaction, the second one is the Dzhyaloshinskii-Moriya interaction, and the third one is the pseudo-dipole interaction. These three components of the exchange Hamiltonian have different symmetry. The scalar interaction conserves both the value of the total spin of the two electrons and its projection on any axis. The pseudo-dipole interaction does not conserve the projection of the total spin (with an exception for the case when the quantization axis is directed along $\vec{\gamma}_{ij}$), but conserves its value. Finally, the Dzyaloshinskii-Moriya interaction does not conserve even the value of the total spin (more precisely, all its matrix elements between the states with the same squared total spin are zero).

### 7) Spin-orbit relaxation mechanisms due to spin diffusion in the impurity band.

As shown in the previous section, in presence of the spin-orbit interaction both tunnelling to empty donors and exchange interaction with other localized electrons may result in a change of the electron spin state. Therefore they can, in principal, bring about the spin relaxation. In this Section, we will consider specific mechanisms of spin-orbit relaxation. We will imply that injection and measurement of the spin polarization is performed in zero magnetic field. In this case the spin kinetics is usually characterized by a single relaxation time $\tau_S$.

Let us start from the relaxation by tunnel hops. Since the binding energies of localized electrons are distributed within the impurity-band energy width, such hops are accompanied by absorption or emission of acoustic phonons. It is a random process. While the electron hops from donor to donor, its spin experiences random rotations (recall that the angle $\gamma(\vec{r}_{ij})$ and the axis of the rotation depend on $\vec{r}_{ij}$). Since $\gamma(\vec{r}_{ij})$ is typically small (of the order of $10^{-2}$ rad), the memory about the initial spin orientation vanishes only after a large number of hops, $N$, when the accumulated rotation angle $\Gamma$ becomes of the order of 1. Using Eq.(7), one can write

$$1 = \Gamma^2 = \sum \langle \gamma_{ij}^2 \rangle = \sum \langle r_{ij}^2 \rangle / L_{SO}^2 = 3 D_h \tau_S / L_{SO}^2 \qquad (9)$$

Thus, the hopping spin relaxation time $\tau_S^h$ can be expressed in terms of the spin-orbit length and the coefficient of hopping diffusion $D_h$ [14]:

$$\tau_S^h = L_{SO}^2 / 3 D_h \qquad (10)$$

Similarly, the isotropic exchange interaction in the ensemble of disordered spins of localized electrons leads to the spin diffusion. Anisotropic corrections to the exchange Hamiltonian result, according to Eq.(7), in a spin rotation through the angle $\gamma_{ij}$ when the spin is transferred between the electrons localized at the centres (*i*) and (*j*). In full analogy with Eq.(10), the spin relaxation time in the case of exchange-dominated relaxation is

$$\tau_S^{ex} = L_{SO}^2 / 3 D_{ex} \qquad (11),$$

where $D_{ex}$ is the coefficient of exchange diffusion. Combining Eqs.(10) and (11), we obtain the expression

$$\tau_S = \left( 1/\tau_S^h + 1/\tau_S^{ex} \right)^{-1} = L_{SO}^2 / 3(D_h + D_{ex}) = L_{SO}^2 / 3 D_S \qquad (12)$$

where $D_S$ is the coefficient of spin diffusion by all the mechanisms. This formula generalizes the conclusions on the relation between transport and spin relaxation, [15, 16, 14], to any kind of spin transport.

The spin-orbit length $L_{SO}$ is determined by the material constants and the binding energy of localized electrons. It is therefore the same for both mechanisms. Thus, the dependence of the spin relaxation time on temperature and impurity concentration is determined solely by the diffusion coefficient.

The temperature and concentration dependence of $D_h$ is known from the theory of hopping conductivity [10]. It decreases exponentially with lowering the donor concentration. Its temperature dependence is also exponential:

$$D_h \propto \exp\left(-\frac{\varepsilon_3}{kT}\right)\exp\left(\frac{\alpha}{N_D^{1/3}a}\right) \tag{13}$$

where $\varepsilon_3 \approx \frac{e^2 N_D^{1/3}}{\varepsilon}$, $N_D$ is the concentration of donors, and $\alpha$ is a number between 1 and 2. With further lowering the temperature into the millikelvin range, a crossover to variable range hopping is possible. In this case, the temperature dependence of the diffusion coefficient is given by the Mott law: $D_h \propto \exp\left(-\left(\frac{T_0}{T}\right)^{1/4}\right)$, where $T_0$ is a characteristic temperature. The concentration dependence of $D_h$ is determined by the exponential decrease of the overlap integral with distance, as well as by the dependence of the efficiency of phonon activation on the energy difference between the two bound electron states. The diffusion goes over donors belonging to the infinite cluster, with the inter-donor distance of the order of the average value ($N_D^{-1/3}$). Therefore, the diffusion coefficient can be written as $D_h \approx \frac{1}{3}kN_D^{-2/3}\tau_W^{-1}$, where $k$ is the compensation degree of the semiconductor (i.e. the number of empty donors per one electron), and the mean waiting time of the tunnel hop is $\tau_W = \tau_{W0}\exp\left(\frac{\varepsilon_3}{kT}+\frac{\alpha}{aN^{1/3}}\right)$. The time parameter $\tau_{W0}$ can be estimated as $\tau_{W0} = \langle w_{ij}^0\rangle^{-1}$ [10], where

$$w_{ij}^0 = \frac{E_D^2 \Delta_{ij}}{\pi d s^5 \hbar^4}\left(\frac{2e^2}{3\varepsilon a}\right)^2 \frac{r_{ij}^2}{a^2}\left[1+\left(\frac{\Delta_{ij}a}{2\hbar s}\right)^2\right]^{-4} \tag{14}$$

Here $E_D$ is the constant of deformation potential, $d$ is the crystal density, $s$ is the sound velocity, $\varepsilon$ is the dielectric constant, and $a$ is the Bohr radius of the donor. Taking the parameters of shallow donors in GaAs, assuming $\alpha = 1.73$ [10] and estimating the characteristic energy difference between different donor-bound states as $\Delta_{ij} \approx \varepsilon_3 = \frac{e^2}{\varepsilon}N_D^{1/3}$, we obtain $\tau_{W0} \approx 5\cdot 10^{-2}(N_D a^3)^3$, that yields $\tau_{W0} \approx 4\cdot 10^{-12}s$ for $N_D=10^{15}$cm$^{-3}$ and $\tau_{W0} \approx 4\cdot 10^{-9}s$ for $N_D=10^{16}$cm$^{-3}$. At T=4.2 K this results in $\tau_W \approx 2\cdot 10^{-3}s$ and $\tau_W \approx 7\cdot 10^{-3}s$, respectively. For a compensated semiconductor, the corresponding diffusion coefficients are the order of $10^{-7}$ cm$^2$/s, in accord with the values, recalculated using the Einstein relation from experimental data on hopping mobility [10]. At weak compensation $D_h$ becomes smaller. With the value of the spin-orbit length for GaAs, $L_{SO} \approx 5\mu$, we obtain $\tau_S^h \approx L_{SO}^2 / D_h \approx 1s$. This result makes us to conclude that the spin-orbit relaxation powered by the hopping diffusion is unable to explain the experimentally measured spin relaxation times n-GaAs at liquid helium temperatures (of the order of $10^{-7}$ s). It is however possible that this mechanism is more effective in quantum wells at low concentrations of electrons, where hopping occurs between states localized at structure imperfections. Beside this, we shall see that hopping may provide the energy relaxation in the spin system of localized electrons.

The spin diffusion by exchange interaction weakly depends on temperature within the range $\langle J \rangle < k_B T < E_B \ln\left(\frac{N_D}{N_C}\right)$, where $N_C$ is the effective density of states in the conduction band. The lower boundary of this range is determined by spin ordering in the system of localized electrons, and the upper boundary - by thermal activation into delocalized states of the conduction band. The compensation dependence is also weak, while $k \ll 1$. The spin diffusion coefficient can be estimated as

$$D_{ex} \approx \frac{1}{3}\langle r_{ij}^2 \rangle \langle J \rangle / \hbar \tag{15}$$

The exchange constant $J$ for hydrogen-like centres is given by the formula [17]:

$$J_{ij} = 0.82\frac{e^2}{\varepsilon a}\left(\frac{r_{ij}}{a}\right)^{5/2} \exp\left(-\frac{2r_{ij}}{a}\right). \tag{16}$$

The exchange constant should be averaged over the infinite cluster; the percolation theory suggests that this can be done by replacing $r_{ij}$ with $\alpha N^{-1/3}/2$, where $\alpha \approx 1.73$ [10]. Substituting the parameters of GaAs, we obtain for $N_D=10^{15}$cm$^{-3}$ $\langle J \rangle \approx 0.1\mu eV$, $D_{ex} \approx 5\cdot 10^{-3} cm^2/s$ and $\tau_S \approx 3\cdot 10^{-6}s$, and for $N_D=10^{16}$cm$^{-3}$ $\langle J \rangle \approx 170\mu eV$, $D_{ex} \approx 1.5 cm^2/s$ and $\tau_S \approx 10^{-7}s$. As we shall see, the model of exchange diffusion demonstrates a very good agreement with the experimental data for GaAs with $N_D$ around $10^{16}$cm$^{-3}$.

The exchange-induced spin relaxation can be alternatively described as a relaxation induced by random fields of anisotropic exchange interaction, in analogy to the Dyakonov-Perel mechanism [7]. The correlation time of this field is

$$\tau_c \approx \langle J \rangle / \hbar \approx \left(3N_D^{2/3} D_{ex}\right)^{-1}, \tag{17}$$

and the corresponding spin precession frequency is $\Omega = \langle J \rangle \langle \gamma_{ij}^2 \rangle^{1/2} / \hbar$.

Applying the motional-averaging formula Eq.(2) then results again in Eq.(14).

It should be noted that the moment-expansion approach to calculation of the spin relaxation rate, used in Ref.[11], is not applicable to the disordered system of donor-bound electrons. The reason is that, because of the exponentially large variation of exchange constants, the means of the second and forth powers of the exchange field, used in that method, are determined by a small number of closely spaced clusters of donors. Therefore, the calculated relaxation times do not characterize the entire ensemble of electrons; for instance, this method gives a wrong concentration dependence of $\tau_s$. The correct approach to calculation of relaxation times in disordered systems with exponential variation of the interaction strength is to use the percolation theory, as done in this Section. The contribution of small clusters should be accounted for separately; this is the subject of the next Section.

*8) Spin-orbit relaxation in small clusters.*

Because of the exponential dependence of the exchange constant on the distance between centres, even a small proportion of donors separated by distances, much less than the average one, may give a sizable contribution into the spin relaxation rate. This idea was first put forward by Liubinskii, Dmitriev and Kachorovskii for the case of hopping relaxation [14]. According to Ref.[14], any electron has a probability to visit small clusters by hopping diffusion. However, as we have shown in the previous Section, the hopping diffusion, at least in bulk semiconductors, is much slower than the exchange spin diffusion. Therefore, the realistic scenario of cluster-dominated spin relaxation is the following: electrons in small clusters are coupled with the rest of donor-bound electrons by isotropic exchange interaction, which determines their spin correlation time $\tau_c \approx \langle J \rangle / \hbar$. However fast is the spin relaxation within the cluster, its contribution into the total spin relaxation rate cannot exceed $(N_{cl}/N_D)/\tau_c$, where $N_{cl}$ is the concentration of such clusters. Since the spin relaxation time by exchange diffusion is $\tau_S^{ex} = L_{SO}^2 / 3D_{ex} = L_{SO}^2 N_D^{-2/3} \tau_c$, the contribution of small clusters may become dominant only if $N_{cl}/N_D > N_D^{-2/3}/L_{SO}^2$. This is the upper estimate of the cluster contribution, corresponding to "black hole" clusters having a very fast spin decay rate. To get more realistic estimates and to find $N_{cl}$, one should consider contributions of two types of relevant clusters: those composed of filled donors, and those where one or more donors are empty.

Let us firstly estimate the contribution of filled clusters, where the spin relaxation is provided by the anisotropic exchange interaction.

The concentration of clusters of closely spaced donors rapidly decreases with increasing the number of donors in the cluster. Therefore we can safely consider only clusters comprising the smallest number of donors, which can provide spin relaxation. It is easy to see that this number is three. Indeed, the exchange Hamiltonian (Eq.(8)) splits the levels of a pair of spins into two groups, a singlet and a triplet. As distinct from the case of purely isotropic exchange, these states are not eigenstates of the total spin, but of an analogous operator composed of electron spins in tilted coordinate frames [12]. But, because of smallness of the angle $\gamma_{ij}$, these states are in fact very close to the eigenstates of the total spin. If, for instance, we place two electrons with the total spin projection +1 on the pair of donors, the probability that they form a singlet state is of the order of $\gamma_{ij}^2 \ll 1$. The probability to occupy the triplet state is $1 - \gamma_{ij}^2 \approx 1$, and, since the levels within the triplet are not split, the total spin projection of the two electrons will not further change and will remain close to 1 forever.

This property of the exchange interaction within a pair of electrons can be better understood if we treat anisotropic terms in Eq.(8) at small $\gamma_{ij}$ as perturbations to the singlet-triplet spectrum formed by the isotropic exchange. As mentioned above, the Dzyaloshinskii-Moriya interaction does not couple states with the same value of the total spin. Therefore, splitting of triplet levels due to this interaction appears as a second-order perturbation. It is easy to find that it is equal to $\gamma_{ij}^2 J/2$. To the opposite, the pseudo-dipole term couples triplet levels directly and results in their splitting in the first order. The value of the energy splitting is exactly equal to one induced by the Dzyaloshinskii-Moriya interaction, but has the opposite sign. As a result, these two contributions cancel each other.

This cancellation of splitting is a result of the specific spectrum of the pair of spins, having only one energy parameter *J*. It does not happen in any other case; in particular, it does not happen for a triad whose spectrum consists of two doublets and a quadruplet [18]. We will not need to perform cumbersome calculations for the case of a triad, because the structure of the spectrum can be guessed using qualitative considerations. The two doublets will not split at all because of the Kramers theorem. The quadruplet, corresponding to the total spin 3/2, will split in two doublets separated by the energy of the order of $\tilde{\gamma}^2 J_{\min}$, where $\tilde{\gamma} = \vec{\gamma}_{12} + \vec{\gamma}_{23} + \vec{\gamma}_{13}$ and $J_{\min}$ is the least of the three exchange constants. This estimate is based on the following considerations. The splitting appears in the second order in the Dzyaloshinskii-Moriya interaction and in the first order in the pseudo-dipole interaction; hence quadratic dependence on $\gamma$. It becomes zero if one of the exchange constants is zero, because in this case we have an open triad, whose Hamiltonian can be written as $J_{12}\vec{S}_1 \cdot \vec{S}_2' + J_{23}\vec{S}_2' \cdot \vec{S}_3''$, where primed and double-primed spin operators are obtained by rotation through angles $\vec{\gamma}_{12}$ and $\vec{\gamma}_{12} + \vec{\gamma}_{23}$ respectively. This Hamiltonian commutes with the operator $\vec{F} = \vec{S}_1 + \vec{S}_2' + \vec{S}_3''$. This operator is an analogue of the total spin, and therefore the multiplets in its spectrum are not split. Finally, it should become zero for the same reason if $\tilde{\gamma} = 0$, because in this case the Hamiltonian can be written as $J_{12}\vec{S}_1 \cdot \vec{S}_2' + J_{23}\vec{S}_2' \cdot \vec{S}_3'' + J_{13}\vec{S}_1 \cdot \vec{S}_3''$ (we assume that the angles are small) and again commutes with $\vec{F}$. The rotation angles in the small closed triad cancel out up to the second order in the case of linear in *k* spin-orbit terms [14]; but in the bulk crystal with cubic spin-orbit terms this does not happen, and one can estimate $\tilde{\gamma} \approx R/L_{SO}$, where *R* is the size of the cluster.

The above consideration suggests that, if a small cluster of three donors with the size *R* is in the quadruplet state (which, for weakly polarized electrons, happens with the probability 0.5), the spin–orbit interaction will make its spin polarization oscillate with the characteristic frequency

$$\Omega_{SO}(R) \approx \left(\frac{R}{L_{SO}}\right)^2 J_{\min}(R)/\hbar \tag{18}$$

The spin relaxation time in the cluster is determined not only by $\Omega_{SO}$, but also by the correlation time of the electron spin, $\tau_c$. If $\Omega_{SO}\tau_c \ll 1$, the spin relaxation time at the cluster is given by the well-known motional-averaging formula (see [19]):

$$\left(\tau_s^{cl}\right)^{-1} \approx \Omega_{SO}^2 \tau_c \tag{19}$$

Otherwise, $\left(\tau_s^{cl}\right)^{-1} \approx \tau_c^{-1}$. In this case, the contribution of clusters into the rate of spin relaxation of the entire electron ensemble is equal to the rate at which the spin polarization is fed into the clusters from surrounding donors ('black hole" regime):

$$w_{bh} = \frac{N_{bh}}{N_D}\tau_c^{-1} \tag{20}$$

Here $N_{bh} \approx \frac{8\pi^2}{9} N_D^2 R_c^6 \exp\left(-\frac{4}{3}\pi N_D R_c^3\right)$ is the concentration of clusters where the largest distance between two donors does not exceed a critical radius $R_c$, defined by the condition

$$\left(\frac{R_c}{L_{SO}}\right)^2 J_{\min}(R_c) = \hbar\tau_c^{-1} = \langle J \rangle.$$

In the regime of short correlation time, the spin relaxation will be governed by clusters of the minimal possible size. At low temperatures, this size is limited by freezing out the clusters where the *largest* of the exchange constants, $J_{\max}$, exceeds $k_B T$. Such clusters remain in the lowest states, which are Kramers doublets, and do not contribute into spin relaxation. Thus, if the correlation time is short, the relaxation will be determined by triads with all the inter-donor distances approximately (with the precision to $\pm a_B$) equal to the "freezing radius" $R_T$, defined by the condition $J(R_T) = k_B T$. The concentration of such clusters can be estimated as $N_T = 4\pi^3 \sqrt{3} N_D^3 R_T^3 a_B^3 \exp\left(-\frac{4}{3}\pi N_D R_T^3\right)$. Their contribution into the overall relaxation rate is

$$w_T = \frac{N_T}{N_D}\left(\frac{R_T}{L_{SO}}\right)^4 \left(\frac{k_B T}{\hbar}\right)^2 \tau_c \tag{21}$$

The crossover to the "black hole" regime occurs when $R_T$ becomes equal to $R_c$, i.e. when $\hbar\tau_c^{-1}$ becomes equal to $\left(\frac{R_T}{L_{SO}}\right)^2 J(R_T)$. At this point (corresponding to $N_D \approx 10^{15} cm^{-3}$), the spin relaxation rate due to exchange interaction in small clusters reaches its maximum, about $10^5$ s$^{-1}$, which is two orders of magnitude faster than the relaxation by exchange diffusion at that concentration. We shall see, however, that this is much slower than the relaxation rate provided in this range of impurity concentrations by the hyperfine interaction.

Unlike the hopping diffusion, which is much slower than its exchange counterpart, phonon-assisted hops within closely-spaced donor clusters can go faster than the spin exchange. But the electron-phonon interaction appears very selective to the spatial and energy separation of the donors, diminishing the number of clusters that actually contribute to the spin relaxation. This is a very important matter for the entire subject, since such hops also provide energy relaxation of the electron spin system, and we shall consider it in detail.

Eq.(14), with which we calculated the hopping probability in the infinite cluster, assumed that the tunnelling matrix element is much smaller than the energy separation of the two donor-bound states. If this condition is not satisfied, $\Delta_{ij}$ in Eq.(14) should be replaced with $\varepsilon_{ij} = \sqrt{\Delta_{ij}^2 + 4I_{ij}^2}$, where $I_{ij} = \frac{4}{3} E_B \frac{r_{ij}}{a} \exp\left(-\frac{r_{ij}}{a}\right)$ is the tunnelling matrix element [10]. This results in the following expression for the waiting time of the hop between donors $i$ and $j$:

$$\tau_w^{-1} = \frac{E_D^2 \varepsilon_{ij} I_{ij}^2}{\pi d s^5 \hbar^4} \left[1 + \left(\frac{\varepsilon_{ij} a}{2\hbar s}\right)^2\right]^{-4} \left[\exp\left(\frac{\varepsilon_{ij}}{k_B T}\right) - 1\right]^{-1} \quad (22)$$

Since $I_{ij} \leq \varepsilon_{ij}/2$, the most frequent hops occur within pairs having $\varepsilon_{ij} \approx \frac{2\hbar s}{a}$, i.e. those interacting with phonons whose wave vectors match the extent of the donor wave function in the k-space. The hopping probability rapidly decreases when $\varepsilon_{ij}$ deviates from this value. The optimal pair should have $\Delta_{ij} \leq 2\hbar s/a$ and $I_{ij} \approx I(r_0) = \hbar s/a$; the latter condition imposes a restriction on the inter-donor distance: $r_{ij} = r_0 \pm a$. For shallow donors in GaAs, $2\hbar s/a \approx 0.12 meV$, yielding $r_0 \approx 4a$. Finally, because usually $2\hbar s/a \approx 0.1 meV < k_B T$, the exponential in the Bose factor can be expanded: $\exp\left(\frac{\varepsilon_{ij}}{k_B T}\right) - 1 \approx \frac{\varepsilon_{ij}}{k_B T}$. The waiting time for the optimal pair then reads:

$$\tau_w^{opt} \approx \left(\frac{E_D^2 k_B T}{16\pi d a_B^2 s^3 \hbar^2}\right)^{-1} \quad (23)$$

For GaAs at liquid helium temperature, $\tau_w^{opt}$ is of the order of 10 ps.

The concentration of optimal pairs is

$$N_{opt} \approx N_D \cdot 4\pi r_0^2 \cdot 2a \cdot (2\hbar s/a)\rho_F, \quad (24)$$

where $\rho_F = kN_D\left(\varepsilon e^2 N_D^{1/3}\right)^{-1}$ is the density of states at the Fermi level in the impurity band [10]. For GaAs with $N_D = 10^{15} cm^{-3}$ and $k$=0.1, $N_{opt}/N_D \approx 4 \cdot 10^{-3}$. The average probability of the hop then writes:

$$\left\langle\left(\tau_w^{opt}\right)^{-1}\right\rangle = \frac{E_D^2 k_B T}{da^2 s^2 \hbar} kN_D r_0^2 \left(e^2 N_D^{1/3}/\varepsilon\right) \quad (25)$$

However, the contribution of such pairs into spin relaxation is suppressed, because the successive hops within the pair are accompanied by spin rotations through the same angle $\gamma_{ij}$ in the opposite directions. As shown by Lyubinskiy [20], the situation changes in a longitudinal magnetic field. The easiest way to understand this effect is to go to the coordinate frame, rotating around the external magnetic field *B* with the Larmor frequency $\omega = \mu_B g B$. In the rotating frame, the external field becomes zero [2], while the spin-orbit fields rotate. Successive hops there and back between the two donors now result in spin rotations about different axes. When the Larmor frequency becomes larger than the inverse waiting time, the correlation between the directions of successive spin turns in the rotating frame is completely lost. As a result, closely spaced pairs start to contribute into the relaxation of all the spin components, including the component along the external field ***B***. Since this component is the same in the rotating and laboratory frames, it follows from the above consideration that the longitudinal magnetic field accelerates the hopping spin relaxation. The resulted spin relaxation rate is equal to $\left\langle \left(\tau_w^{opt}\right)^{-1} \right\rangle \cdot \left(L_{SO}/r_0\right)^2$, which amounts to approximately $10^5$ s$^{-1}$ for GaAs with $N_D = 10^{15} cm^{-3}$ and *k*=0.1.

The results of this Section can be summarized as follows. Relaxation in small clusters, either by phonon-assisted tunnelling, or by anisotropic exchange interaction, limits the spin relaxation time at the level of, approximately, $10^{-5}$ s. At donor concentration around $10^{15}$ cm$^{-3}$ this time is shorter than one due to exchange diffusion. But, as shown in the next Section, spin relaxation by nuclei is much faster in this doping range.

### 9) *Relaxation of the electron spins by nuclei.*

The localized electron is coupled with a large number of nuclear spins within its orbit by the Fermi contact interaction, proportional to the scalar product of the electron and nuclear spins and to the squared electron wave function at the location of the nucleus [3, 21]. The hyperfine interaction can be expressed in terms of the effective nuclear magnetic field $B_N$ applied to the electron spin. This field can be as strong as a few Tesla, if nuclear spins are polarized. If they are not polarized, there is still some fluctuation nuclear field due to incomplete compensation of fields produced by randomly directed nuclear spins. This fluctuation field is described by the Gaussian statistics. Its root-mean-square value can be estimated as

$$\left\langle B_{Nf}^2 \right\rangle^{1/2} = B_{N\max} / \sqrt{N_N} \qquad (26)$$

where $B_{N\max}$ is the maximum value of the nuclear field, corresponding to fully polarized nuclear spins, and $N_N$ is the number of nuclei in the localization volume of the electron. Typically, $N_N$ is of the order of $10^5$. A calculation for the case of a shallow donor in GaAs gives $\left\langle B_{Nf}^2 \right\rangle^{1/2} = 54G$ [22]. Because of the smallness of magnetic moments of nuclei as compared to that of the electron, nuclear spins evolve on much longer time scales than electron spins. For this reason, nuclear fields can always be considered quasi-stationary. The regimes of spin relaxation in this situation are governed by the electron correlation time $\tau_c$. If $\mu_B g \langle B_{Nf}^2 \rangle^{1/2} \tau_c > 1$, all the components of the electron spin, perpendicular to the local nuclear field, will disappear on average during the period of spin precession in the nuclear field (typically, a few nanoseconds). The remaining polarization, which amounts to 1/3 of the initial value, relaxes during much longer time determined by the nuclear spin dynamics [23, 24]. If $\frac{\mu_B g}{\hbar} \langle B_{Nf}^2 \rangle^{1/2} \tau_c \ll 1$ (the regime of short correlation time), the time of electron spin relaxation by nuclei is given by the motional-averaging formula:

$$\tau_{sN}^{-1} = \left( \frac{\mu_B g}{\hbar} \right)^2 \langle B_{Nf}^2 \rangle \tau_c \qquad (27)$$

The most powerful mechanism limiting the electron correlation time in the impurity band is the exchange-induced spin diffusion. Therefore, $\tau_c \approx \left( N_D^{2/3} D_{ex} \right)^{-1}$, and one can expect $\tau_{sN}$ to increase rapidly with increasing the donor concentration, starting from several nanoseconds in the most lightly doped crystals (see Fig.3)

10) **Interaction with free electrons.**

Since at low concentration of donors localized electrons are well isolated from each other, even a small concentration of free electrons in the conduction band may strongly affect the spin correlation time of bound electrons. The probability of spin exchange between a donor-bound electron and a free electron in the conduction band can be calculated using the results of the theory of electron scattering by atomic hydrogen [25]. It is determined by the difference of phase shifts for the triplet and singlet scattering. If, for example, the bound electron is in the spin-up state, and the free electron is in the spin-down state, the two-electron spin state can be written as a superposition of the states with the total spin *I* equal to 0 and 1:

$$\uparrow\downarrow = \frac{1}{\sqrt{2}} \left[ \frac{1}{\sqrt{2}} (\uparrow\downarrow - \downarrow\uparrow) + \frac{1}{\sqrt{2}} (\uparrow\downarrow + \downarrow\uparrow) \right] = \frac{1}{\sqrt{2}} [(1, m=0) + (0, m=0)], \qquad (28)$$

which, after scattering, transforms into

$$\frac{1}{\sqrt{2}} \left[ \frac{1}{\sqrt{2}} f_s (\uparrow\downarrow - \downarrow\uparrow) + f_t \frac{1}{\sqrt{2}} (\uparrow\downarrow + \downarrow\uparrow) \right] = \frac{1}{2} [(f_s + f_t)\uparrow\downarrow + (f_s - f_t)\downarrow\uparrow] \qquad (29)$$

where m denotes the spin projection. In the limit of low kinetic energy of electrons, scattering amplitudes for *I*=0 (singlet) and *I*=1(triplet) states are angular-independent and equal to [18]:

$$\begin{aligned} f_s &= k^{-1} \sin(\eta_{0s}) e^{i\eta_{0s}} \\ f_t &= k^{-1} \sin(\eta_{0t}) e^{i\eta_{0t}} \end{aligned} \qquad (30)$$

where $\eta_{0s}$ and $\eta_{0t}$ are zeroth-order phase shifts. The spin-flip scattering cross-section now writes

$$\sigma_{sf} = \frac{4\pi}{k^2}|f_s - f_t|^2 = \frac{\pi}{k^2}\sin^2(\eta_{0s} - \eta_{0t}) \tag{31}$$

The phase shifts have been calculated numerically and can be found in the literature on atomic collisions. Using the dependence of $\eta_{0s} - \eta_{0t}$ on $ka_B$ from Fig.1 of Ref. [25], one may propose an approximation formula:

$$\sigma_{sf}(k) \approx \frac{20.6\pi a_B^2}{1 + (3.9ka_B)^3} \tag{32}$$

reproducing the numerical results with the precision better than 0.02 for $ka_B \leq 0.7$. The probability for a bound electron to flip its spin as a result of collision with a free one is

$$w_{sf}^b = \langle \sigma_{sf}(k)\hbar k n_c / m \rangle = \frac{n_c}{N_C}\frac{\hbar}{\pi m a_B^2} Q(b) \tag{33}$$

where $Q(b) = \int_0^\infty \frac{20.6 x^3}{1 + (3.9x)^3}\exp(-x^2/b)$, $b = \frac{k_B T}{E_B}$, $n_c$ is the concentration of free electrons, and $N_C$ is the effective density of states in the conduction band. We assume that $n_c \ll N_C$ and therefore use the Boltzmann statistics. It is worth noting that at $N_D > 10^{16}$ cm$^{-3}$ (for GaAs) $\sigma_{sf}$ may exceed the mean squared distance between adjacent donors; at such impurity concentrations, the model of independent scatterings fails, and Eq.(33) can be used only for rough estimations. Equation (33) is applicable when the thermal energy is less than the Bohr energy of donors; if this condition is satisfied, the function $Q$ can be, with a good precision, approximated by the formula

$$Q(b) \approx 0.192\left[\sqrt{1 + (10b)^2} - 1\right]. \tag{34}$$

In thermal equilibrium, the electron concentration in the conduction band is $n_c \approx N_C \exp(-E_B/k_B T)$, corresponding to $n_c \approx 6 \cdot 10^{10} cm^{-3}$ at T=4.2K, and $n_c \approx 4 \cdot 10^5 cm^{-3}$ at T=2K. Using Eq.(33), we obtain $w_{sf}^b \approx 3 \cdot 10^9 s^{-1}$ at T=10K, $w_{sf}^b \approx 3 \cdot 10^5 s^{-1}$ at T=4.2K, and $w_{sf}^b \approx 0.02 s^{-1}$ at T=2K. Under optical excitation $n_c$ can be, of course, much higher; it depends on the excitation wavelength and intensity, doping and temperature and may vary very strongly from experiment to experiment.

The probability for the free electron to flip its spin in a collision with a donor-bound electron is equal to

$$w_{sf}^i = \frac{N_D}{N_C}\frac{\hbar}{\pi m a_B^2} Q(b). \tag{35}$$

It is as large as, approximately, $10^{10}$ s$^{-1}$ already at $N_D = 10^{14}$ cm$^{-2}$. This means that exchange scattering by neutral donors is the main spin relaxation mechanism for conduction-band electrons at low temperatures, and their mean spin is equal to the mean spin of localized electrons with a good precision [26]. Therefore, localized and free electrons form a spin system characterized by common relaxation times.

## 11) *The influence of longitudinal magnetic fields on spin relaxation.*

Spin relaxation in zero or very weak magnetic field requires only breaking the angular-momentum conservation (which is realized by hyperfine and/or spin-orbit interactions). In longitudinal (i.e. parallel to the mean spin) magnetic field of considerable strength, changing the mean spin of electrons is accompanied by changing their energy, which should be eventually dissipated into the crystal lattice. The energy relaxation of the electron spin system can go in one or two steps, depending on the strength of the magnetic field applied.

In strong fields, $B >> \langle J \rangle / \mu_B g$, the energy $\mu_B g B$, released in the spin-flip transition, cannot be absorbed by the spin system, and the transition should be accompanied with absorption/emission of a phonon. The phonon-assisted spin relaxation in strong magnetic fields has been a subject of many theoretical works [9, 27]. As collective spin interactions in the impurity band are less important for this process, we will not consider it here.

In weak-to-moderate fields, $B \leq \langle J \rangle / \mu_B g$, the spin-flip transition can go without phonon assistance, the energy being temporarily stored within the spin system in the form of exchange energy. Its dissipation into the lattice goes independently, by phonon-assisted transitions within the energy spectrum of the spin system, broadened by the exchange interaction. The issues related to the energy relaxation of the electron spin system in magnetic fields will be discussed in the next Section

12.

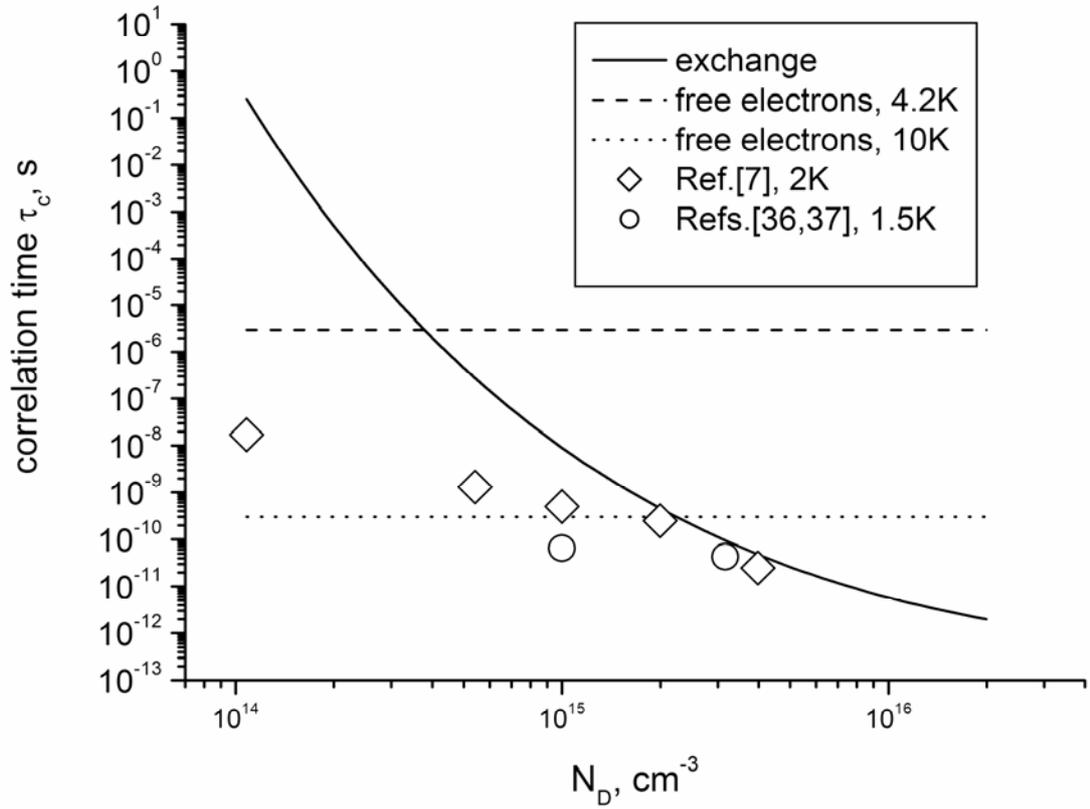

Fig.1. Spin correlation time determined by exchange diffusion and interaction with free electrons vs donor concentration, calculated using the parameters of GaAs. Symbols: experimental values, determined from suppression of spin relaxation by longitudinal magnetic fields.

Here, we will concentrate on the magnetic-field dependence of the relaxation rate of the non-equilibrium angular momentum. In this view, it is worth to recall general expressions for spin relaxation induced by random magnetic fields [19]. If the correlation time of the random field is shorter than the period of the electron spin precession in this field, then the spin relaxation rate in zero external field is given by the motional-averaging formula (Eqs.(2), (19), (27)). Applying a constant longitudinal field $B$ slows down the spin relaxation by the factor $\frac{1}{1+(\omega\tau_c)^2}$, where $\omega = \mu_B g B / \hbar$. The characteristic field that diminishes the spin relaxation rate two times is then equal to $B_{1/2} = \frac{\hbar}{\mu_B g \tau_c}$. This expression is very general, but not universal. It is true when the random field is characterized by a single correlation time. This is correct in the case of exchange-induced relaxation, with $\tau_c \approx \left(N_D^{2/3} D_{ex}\right)^{-1}$. In the case of hopping relaxation, however, there are two very different time parameters of equal importance: the duration of a single hopping transition $\tau_h$ and the waiting time $\tau_w$. Since the spin-orbit field affects the spin during the hopping transition, it is the time $\tau_h$ that enters the suppression factor. The hopping transition takes the time of the order of the inverse frequency of phonons assisting the tunnelling. As a result, in the case of hopping the suppression field is

$$B_{1/2} = \frac{\hbar}{\mu_B g \tau_h} \approx \langle \Delta_{ij} \rangle / \mu_B g \quad (36)$$

For GaAs with $N_D=10^{15}$cm$^{-3}$ $\tau_h$ is of the order of 1ps, and this field exceeds 10 Tesla. The much longer waiting time $\tau_w$ may also show up, in a rather unexpected way, in the field dependence of the spin relaxation rate. As shown by Lyubinskiy [20], relatively weak magnetic fields, of the order of $\frac{1}{\mu_B g \tau_w}$, make the spin relaxation *faster*. This unusual effect, resulted from the contribution of closely spaced pairs of donors into the spin relaxation, has been discussed in Section 8.

Spin relaxation can be also affected by magnetic localization of electrons. Magnetic fields are known to strongly suppress the hopping conductivity by diminishing the overlap of wave functions of localized electrons [10]. In magnetic fields up to several Tesla, the hopping conductivity decreases as an exponential function of the squared magnetic field. As both the hopping probability and the exchange constant are proportional to the same exponential factor $\exp(-2r_{ij}/a_B)$, the magnetic-field dependence known from the theory of hopping conductivity [10] can be universally applied to spin diffusion in the impurity band:

$$D_S(B) = D_S(0) \exp\left(-0.04 \frac{a_B e^2}{N_D c^2 \hbar^2} B^2\right) \quad (37)$$

In case of GaAs, the exponent reaches 1 at $B=1$T for $N_D=10^{15}$ cm$^{-3}$, and at $B=3$T for $N_D=10^{15}$ cm$^{-3}$. These fields are of the same order as $B_{1/2} = \frac{\hbar}{\mu_B g \tau_c}$. Having in mind that there are several contributions to the spin relaxation rate: exchange diffusion, hyperfine interaction, hopping and exchange in small clusters, depending differently on $\tau_c \propto D_S^{-1}$, one can expect, generally, a complex pattern of the magnetic-field influence on spin relaxation.

*12) Energy relaxation of electron spins in longitudinal magnetic field.*
The anisotropic exchange interaction provides relaxation of the non-equilibrium polarization of electron spins, but does not provide transfer of their energy to the crystal lattice. The same is true for the relaxation due to the hyperfine interaction: in that case, electron spins are coupled only with the nuclear spin system having a very small heat capacity and very long energy relaxation time [3].

For that reason, spin relaxation in the system of localized electrons is, generally, characterized by two times rather than a single time $\tau_s$: the relaxation time of the non-equilibrium angular momentum, $T_2$ (in zero magnetic field, $T_2=\tau_s$), and the energy relaxation time, $T_1$. If the condition $T_2 \ll T_1$ is satisfied, the spin system can be characterized by a spin temperature $\theta$, which can be different from the lattice temperature $T$ (for an introduction into the concept of spin temperature, see Ref.[2]). In that case, optical spin orientation in the longitudinal magnetic field, when a change of the polarization of spins is accompanied by changing their energy, should be interpreted as cooling of the spin system. The spin cooling is well known for nuclear spin systems; now we should find out whether or not it may occur for electrons.

Let us estimate the energy relaxation time $T_1$. The energy transfer between the spin system and the lattice may result from phonon-assisted spin flips of electrons. But the spin-phonon scattering time of a single localized electron at liquid-helium temperatures is about 0.1 s or longer [9]. The spin relaxation by phonon-assisted hopping is also ineffective, as shown above. However, hopping can provide energy relaxation of the electron spin system even without spin flips. Indeed, when an electron hops from donor to donor, the constants of its exchange interaction with other electrons change, due to their exponential dependence on distance, by the values of the order of themselves. As a result, the energy of the spin system changes by the value of the order of the mean exchange energy per one electron. The total number of hops in unit time is determined by optimal pairs (see Section 8). The waiting time for a hop in such a pair, $\tau_{opt}^w$, given by Eq.(23), is of the order of 10ps. If the correlation time is shorter, the hopping contribution into the energy relaxation rate, $1/T_{1h}$, is of the order of $(\tau_{opt}^w)^{-1} N_{opt} / N_D$, where the concentration of optimal pairs is given by Eq.(24). If $\tau_{opt}^w$ is shorter than $\tau_c$, $1/T_{1h}$ is determined by the energy transfer to optimal pairs from other electrons and can be estimated as $(\tau_c)^{-1} N_{opt} / N_D$. An approximation formula,

$$T_{1h} \approx (\tau_c + \tau_{opt}^w) N_D / N_{opt} \tag{38}$$

can be used to calculate $T_{1h}$ in both regimes. In GaAs, the crossover between the two regimes occurs at donor concentrations near $10^{16}$ cm$^{-3}$, where $T_{1h} \approx 10^{-9} s$. With decreasing concentration below $N_D = 10^{16}$ cm$^{-3}$, $T_{1h}$ becomes longer, following the increase of $\tau_c$. Eventually it becomes as long as several microseconds at $N_D$ around $10^{15}$ cm$^{-3}$, exceeding the spin relaxation time, which in that range of concentrations is limited by the hyperfine interaction.

Another possible mechanism of energy relaxation is thermal activation of localized electrons into the conduction band. One can expect that in the temperature range where hopping conductivity dominates, activation is a more rare event than a tunnel hop. A quantitative estimate confirms this. The probability of thermal activation is given by the principle of detailed equilibrium:

$$w_a = \frac{1}{2} N_c \sigma_c v_t \exp(-E_B / k_B T) \tag{39}$$

where $N_c = 2\dfrac{(2\pi m k_B T)^{3/2}}{(2\pi\hbar)^3}$ is the effective density of states in the conduction band,

$\sigma_c = \dfrac{8m^3 E_D^2}{3\hbar^4 d}\left(\dfrac{e^2}{\varepsilon k_B T}\right)^3$ is the cross-section of capture of a free electron to the donor-bound state,

$v_t = \sqrt{\dfrac{8k_B T}{\pi m}}$ is the thermal velocity [28]. For GaAs, $w_a$ is 200 s$^{-1}$ at $T$=4.2 K, and 10$^{-6}$ s$^{-1}$ at $T$=2 K.

The interaction with free electrons, considered above (Section 10), can also provide energy relaxation. Free electrons receive energy from localized ones at the rate $\langle\varepsilon_{ex}\rangle w_{sf}^i$ and give it up to the lattice at the rate $k_B T_e/\tau_e$, where $T_e$ is the kinetic temperature of free electrons, $\tau_e$ is their energy relaxation time by phonon emission, and $\langle\varepsilon_{ex}\rangle$ is the mean exchange energy of localized electrons. The time $T_1^i$, characterizing the energy relaxation via free electrons is determined by these two successive processes:

$$T_1^i = \left(w_{sf}^b\right)^{-1} + \dfrac{N_D\langle\varepsilon_{ex}\rangle}{n_c k_B T_e}\tau_e, \qquad (40)$$

At liquid-helium temperatures, $\tau_e$ is of the order of nanoseconds. Using the results of Section 10, one can estimate that, if the electron concentration in the conduction band is determined by thermal activation, $T_1^i$ is of the order of 10$^{-4}$ s or longer. Presence of optically pumped electrons can, however, make it much shorter.

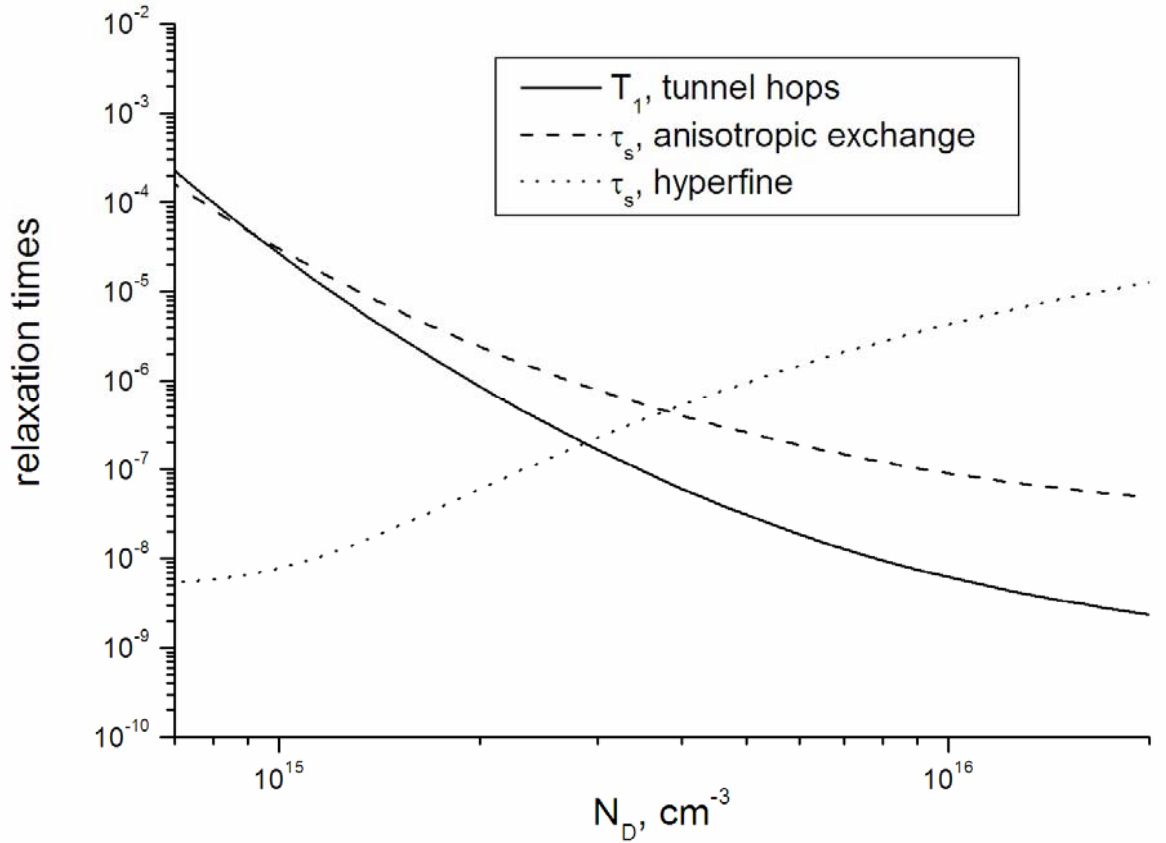

Fig.2. Energy relaxation time $T_1$ determined by hopping in optimal pairs, compared with the spin relaxation time determined by the anisotropic exchange interaction and hyperfine interaction. Calculations are performed with the parameters of GaAs, T=4.2K.

Summarizing the above paragraphs, we can conclude that the energy relaxation of the spin system of bound electrons is provided mainly by tunnel hops in optimal pairs of donors. At high donor concentrations, the energy relaxation is faster than the spin relaxation determined by the anisotropic exchange interaction. With lowering concentration, the energy relaxation becomes less effective because of increasing isolation of donors from each other, which slows down the energy transfer to optimal pairs (see Fig.2). At the same time, the spin relaxation time gets shorter due to the hyperfine relaxation. As a result, the inequality between relaxation times of angular momentum and energy becomes weaker and even reverses. This also may happen in the case when the number of empty donor states is very low – for example, in heterostructures where additional electrons may come from barriers. In all these situations, the decay of spin polarization in a longitudinal magnetic field will be limited by the energy relaxation of the spin system. The spin dynamics in this regime will be considered in the next Section.

*13) Spin dynamics under cooling of the electron spin system.*

If the non-equilibrium mean spin relaxes faster than the energy, the system of interacting spins can be characterized by a spin temperature, which may be different from the lattice temperature. The nuclear spin system of a semiconductor [3] is a well-known example. Let us now obtain differential equations for the mean spin and spin temperature of the system of donor-bound electrons under optical pumping.

The influx of mean spin due to optical spin orientation in n-type semiconductors is equal to $\frac{S_0 - S}{\tau_j}$, where $S_0$ is the mean spin of optically excited electrons, $S$ is the mean spin of resident electrons, and $\tau_j$ is the characteristic time of replacement of a resident electron by a photo-excited one ($\tau_j$ is inversely proportional to the pumping intensity) [29]. Here we consider only the spin components along the external magnetic field.

In the longitudinal magnetic field, this spin influx is accompanied by an energy influx, equal to $\mu_B g B \frac{S_0 - S}{\tau_j}$ (per one electron). The rate of change of the reciprocal spin temperature $\beta = \frac{1}{k_B \theta}$ due to this energy influx equals $\mu_B g B \frac{S_0 - S}{\tau_j} \left( \frac{\partial U}{\partial \beta} \right)^{-1}$, where $U$ is the total energy of the spin system; $U = Tr(\hat{\rho}\hat{H})$, where $\hat{\rho}$ is the spin density matrix, $\hat{H}$ is the Hamiltonian of the spin system. Taking the derivative of energy by the reciprocal temperature, we obtain:

$$\frac{\partial U}{\partial \beta} = \frac{\partial}{\partial \beta} Tr(\hat{\rho}\hat{H}) = \frac{\partial}{\partial \beta} \frac{\sum_i \varepsilon_i \exp(-\beta \varepsilon_i)}{\sum_i \exp(-\beta \varepsilon_i)} =$$

$$\frac{\sum_i \varepsilon^2_i \exp(-\beta \varepsilon_i)}{\sum_i \exp(-\beta \varepsilon_i)} - \frac{\left( \sum_i \varepsilon_i \exp(-\beta \varepsilon_i) \right)^2}{\left( \sum_i \exp(-\beta \varepsilon_i) \right)^2} = \langle \varepsilon^2 \rangle - \langle \varepsilon \rangle^2 = \Delta \varepsilon^2$$

(41)

where $\varepsilon_i$ is the energy of $i$-th spin state.

Since the isotropic exchange and Zeeman interactions commute, and the anisotropic correction are small, one can write:

$$\frac{\partial U}{\partial \beta} = \Delta \varepsilon_z^2 + \Delta \varepsilon_{ex}^2$$

(42)

If the spin ensemble is weakly polarized, $\Delta \varepsilon_{ex}^2 \approx \langle J^2 \rangle$, and $\Delta \varepsilon_z^2 = \langle (\mu_B g B s_z)^2 \rangle \approx \frac{1}{4}(\mu_B g B)^2$.

Optical pumping also results in an increase of the exchange energy due to disruption of spin-spin correlations. In the high-temperature approximation, the mean exchange energy of a localized electron equals $\Delta \varepsilon_{ex}^2 \beta$. Replacement of resident localized electrons with photo-excited ones having zero exchange energy results in the energy influx $\frac{\Delta \varepsilon_{ex}^2 \beta}{\tau_j}$.

Finally, the reciprocal spin temperature relaxes to the reciprocal lattice temperature with the time $T_1$, and the mean spin $S$ relaxes to its quasi-equilibrium value, corresponding to the reciprocal spin temperature $\beta$, $S_\beta \approx \dfrac{\beta}{4}\mu_B g B$, with the time $T_2$.

Now we can write the differential equations for $\beta$ and $S$:

$$\begin{cases} \dot{\beta} = \dfrac{(\mu_B g B + \langle J \rangle S)}{\mu_B^2 g^2 B^2/4 + \Delta\varepsilon_{ex}^2}\dfrac{S_0 - S}{\tau_j} - \dfrac{\Delta\varepsilon_{ex}^2}{\mu_B^2 g^2 B^2/4 + \Delta\varepsilon_{ex}^2}\dfrac{\beta}{\tau_j} - \dfrac{\beta - 1/k_B T}{T_1} \\ \dot{S} = \dfrac{S_0 - S}{\tau_j} - \dfrac{1}{T_2}(S - S_\beta) \end{cases} \quad (43)$$

It is convenient to rewrite these equations in terms of the current value of the mean spin, $S$, and its quasi-equilibrium value, $S_\beta$:

$$\begin{cases} \dot{S}_\beta = f(B)\left(1 + \dfrac{\langle J \rangle}{\mu_B g B}S\right)\dfrac{S_0 - S}{\tau_j} - (1 - f(B))\dfrac{S_\beta}{\tau_j} - \dfrac{S_\beta - S_L}{T_1} \\ \dot{S} = \dfrac{S_0 - S}{\tau_j} - \dfrac{1}{T_2}(S - S_\beta) \end{cases} \quad (44)$$

where $f(B) = \dfrac{(\mu_B g B)^2}{(\mu_B g B)^2 + 4\Delta\varepsilon_{ex}^2} \approx \dfrac{(\mu_B g B)^2}{(\mu_B g B)^2 + 3\langle J \rangle^2/4}$, and $S_L \approx \dfrac{1}{4k_B T}\mu_B g B$ is the equilibrium value of the mean spin at the lattice temperature.

Under the constant-wave excitation, Eq.(44) yields the following expressions for the quasi-equilibrium mean spin, $S_\beta$, and the non-equilibrium part of the mean spin, $S - S_\beta$:

$$\begin{cases} S_\beta = \dfrac{f(B)S_0 + (T_2 + \tau_j)T_1^{-1} S_L}{1 + (1 - f(B))T_2\tau_j^{-1} + (T_2 + \tau_j)T_1^{-1}} \\ S - S_\beta = \dfrac{T_2}{\tau_j}\dfrac{(1 - f(B) + \tau_j T_1^{-1})S_0 - \tau_j T_1^{-1} S_L}{1 + (1 - f(B))T_2\tau_j^{-1} + (T_2 + \tau_j)T_1^{-1}} \end{cases} \quad (45)$$

With increasing the pump intensity, the mean spin changes from the $S \approx S_L$ at $\tau_j \gg T_1$ to $S \approx S_0$ at $\tau_j \ll T_2$. Under moderate pump intensity, when $T_2 \ll \tau_j \ll T_1$, Eq.(45) simplify:

$$\begin{cases} S - S_\beta \approx \dfrac{T_2}{\tau_j}S_0 \\ S_\beta \approx f(B)S_0 + \dfrac{\tau_j}{T_1}S_L \end{cases} \quad (46)$$

The decay of spin polarization after pumping by a pulse of circularly polarized light is described by Eq.(44) at $\tau_j = \infty$:

$$\begin{cases} \dot{S}_\beta = -\dfrac{S_\beta - S_L}{T_1} \\ \dot{S} = -\dfrac{1}{T_2}(S - S_\beta) \end{cases}$$

(47)

The increase of the magnetic field results in changing the quasi-equilibrium part of the mean spin from a small value $\dfrac{\tau_j}{T_1} S_L$, determined by heating the spin system by optical pumping, to the value close to the mean spin of photo-excited electrons, $S_0$. In weak magnetic fields, $B << \dfrac{2\Delta\varepsilon_{ex}}{\mu_B g}\sqrt{\dfrac{T_2}{\tau_j}}$, the spin polarization is purely non-equilibrium, and an exponential decay with the time $T_2$ should be observed. In intermediate fields, of the order of $\dfrac{2\Delta\varepsilon_{ex}}{\mu_B g}\sqrt{\dfrac{T_2}{\tau_j}}$, a two-exponential decay is expected: first the decay of the non-equilibrium spin with the time $T_2$, and then diminishing of the quasi-equilibrium part due to relaxation of the spin temperature with the time $T_1$. Finally, in moderately strong fields, $\dfrac{2\Delta\varepsilon_{ex}}{\mu_B g}\sqrt{\dfrac{T_2}{\tau_j}} < B < \dfrac{\Delta\varepsilon_{ex}}{\mu_B g}$, only the quasi-equilibrium polarization remains, decaying with the time $T_1$. It is worth noting that all these changes occur in magnetic fields too small to affect $T_2$, which increases in the characteristic field $B_{1/2} = \dfrac{\hbar}{\mu_B g \tau_c} \approx \dfrac{\Delta\varepsilon_{ex}}{\mu_B g}$.

One can see that, in a general case, there are two decay times, $T_1$ and $T_2$. The observed relaxation pattern is determined largely by the excitation conditions: magnetic field in which the electrons are excited, excitation intensity and its spectral position, which may strongly affect $T_1$ by changing the concentration of photoexcited conduction-band electrons or excitons.

*14) Possible pitfalls for experiments on spin relaxation.*

Spin relaxation in n-type semiconductors is studied with a variety of experimental techniques, mainly using optical orientation of electron spins. Recently, spin noise spectroscopy, developed earlier [30] for atomic gases, has been modified and applied to semiconductors [31]. The latter method has an apparent advantage of not perturbing the spin system studied, while none of the more traditional approaches is fully free of this shortcoming.

There are several ways by which optical pumping affects spin relaxation. Most obvious are listed below:

1. The spin lifetime is limited by the recombination time of electrons with photoexcited holes. This has been understood since classical works of Dyakonov and Perel [29], and normally precautions are taken to avoid this effect. It is easily eliminated by determining $\tau_s$ from a cut-off of the dependence of the Hanle curve width on pumping at zero pump intensity [7] or by measuring the spin dynamics after the hole recombination time in experiments with pulsed excitation [31]. The same measures remove the effects of the exchange scattering by holes (Bir-Aronov-Pikus spin relaxation mechanism).

2. A population of delocalised electrons is created in the conduction band, which remains there after recombination of holes with localized electrons. The concentration of free electrons in bulk crystals decreases back to the equilibrium value during the capture time $\tau_t = (kN_D\sigma_c v_t)^{-1}$, which for GaAs at liquid-helium temperatures amounts to about 1 ns.

The situation can be different in heterostructures where re-charging effects can take place. This was demonstrated in Ref.[22] by changing the excitation wavelength. Under the illumination just above the bandgap, additional electrons came to the GaAs layer ($N_D=10^{14}$cm$^{-3}$) from AlGaAs barriers. There were not enough donor-bound states to accommodate them, and a population of free electrons was created. As a result of exchange scattering, the correlation time of donor-bound electrons was reduced to approximately $10^{-10}$s; according to Eqs.(33) and (34), this requires about $7 \cdot 10^{13}$ free electrons per cubic centimetre. At such a short correlation time, relaxation by nuclei was suppressed; Eq.(27) gives the spin relaxation time of about 300ns, which was indeed measured using the Hanle effect under resonant excitation near the band edge with a tuneable Ti-Sapphire laser [22]. Under illumination with much higher photon energy, additional electrons were removed from the GaAs layer, and the spin relaxation time dropped down to 5ns, which corresponds to the long-correlation-time regime of hyperfine relaxation, typical for isolated donors.

It should be noted that the effect of illumination may be different depending on the structure design, wavelength and intensity of light. Often charge carriers are transferred not from, but into the studied layers [32]. In lightly doped crystals, spin density of photoexcited electrons may exceed that of localized ones and mask their spin dynamics.

3. Excitation light can heat up delocalised electrons and/or change the spin temperature of localized ones, as described in Section 13. This can affect, first of all, measurements of the spin correlation time using suppression of spin relaxation by longitudinal magnetic fields. The increased relaxation time observed in a magnetic field may be in fact the energy relaxation time $T_1$, which determines the spin dynamics of the cooled spin system.

15) *Experimental studies of spin relaxation of localized electrons in n-type semiconductors: Past, present and future.*

Extended (up to 30 ns) spin relaxation times in lightly doped n-type crystals of GaAs were observed already at early stages of research on optical spin orientation by Weisbuch [33]. In late 90-es, Dzhioev et al [34] and Kikkawa and Awschalom [35] reported measuring $\tau_s = 42ns$ at $N_D = 4 \cdot 10^{15}$ cm$^{-3}$ and 130 ns at $N_D = 10^{16}$ cm$^{-3}$, correspondingly. The most comprehensive, to this date, study of spin relaxation in bulk n-type semiconductors was performed by Dzhioev et al [7]. Spin relaxation times were measured using the conventional Hanle effect (depolarisation of photoluminescence by a transversal magnetic field) in bulk GaAs crystals with $N_D$ spanning the range from $10^{14}$ to $10^{17}$ cm$^{-3}$. For most concentrations, the measurements were performed at two temperatures, 4.2K and 2K, not showing a significant difference in $\tau_s$ between these two temperatures. For a few concentrations below $N_D = 5 \cdot 10^{15}$ cm$^{-3}$, dependences of the polarization of photoluminescence on the longitudinal magnetic fields were measured and used to determine the correlation time $\tau_c$ (see Section 11). Later, the results for several other samples within the same concentration range were reported, measured with a time-resolved photoluminescence technique [36, 37] and using the Hanle effect detected with the photoinduced Kerr rotation [38]. In Refs.[36, 37], spin relaxation time was measured as a function of temperature and longitudinal magnetic field. The experimentally measured spin relaxation times from Refs.[ 7, 35-38] are plotted in Fig.3 against donor concentration in the range below $N_D = 2 \cdot 10^{16}$ cm$^{-3}$ (above this concentration, electrons in GaAs are delocalized). The theoretical curves are calculated for the hyperfine and anisotropic exchange mechanisms. Since there is no fitting parameters in the theory, the agreement with the majority of the experimental data looks remarkable. Still, some discrepancies are seen. Firstly, the relaxation times for $N_D = 4 \cdot 10^{15}$ cm$^{-3}$ and $N_D = 10^{16}$ cm$^{-3}$, reported in Ref.[38], are much longer than those measured by other groups. Secondly, experiments in lightly-doped samples show very large scattering and generally longer times than predicted by theory. Thirdly, the experiments of Refs.[36, 37] demonstrate rather strong temperature dependence of the spin relaxation time – in contrast to the data of Ref.[7] revealing practically no difference between $\tau_s$ at 2K and 4.2K.

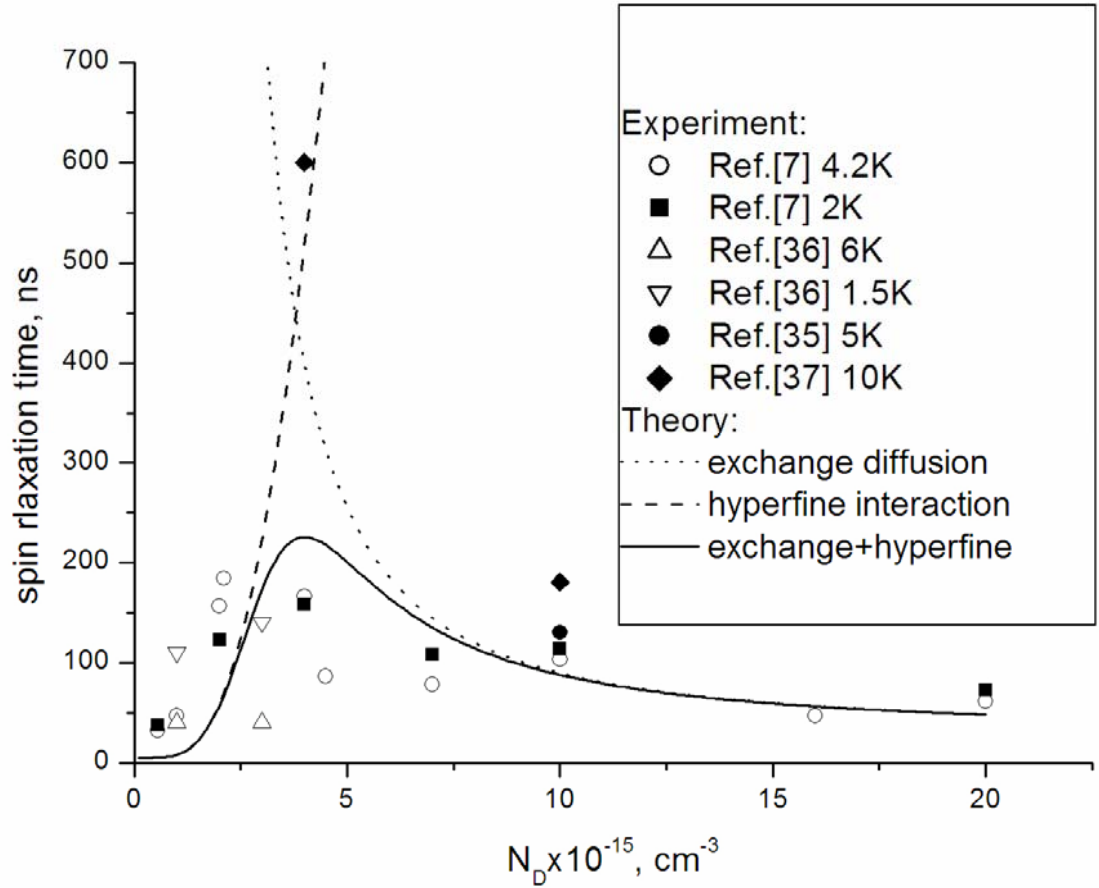

Fig.3. Experimental data on spin relaxation time in GaAs in zero or weak magnetic field vs donor concentration. Lines present the theory taking into account two most powerful relaxation mechanisms: by hyperfine interaction and spin-orbit interaction in course of exchange diffusion.

These features are probably related to the behaviour of the spin correlation time that, in the motion-narrowing regime, determines the spin decay rate via the hyperfine interaction. This time was measured from the dependences of the electron mean spin or spin relaxation time on the longitudinal magnetic field [7, 36, 37]. As seen from Fig.1, in the low-doping range experimentally measured values of $\tau_c$ are systematically shorter than theoretical ones, determined by the exchange diffusion. In addition, a non-monotonous magnetic-field dependence of the spin relaxation time was observed in a GaAs sample with $N_D = 10^{15} cm^{-3}$ [37]. The origin of these effects is not quite clear, but most likely it is interaction with free electrons. The concentration of the latter in those experiments might have exceeded the thermally equilibrium value due to either optical pumping or injection from barriers in heterostructure samples. This phenomenon, qualitatively demonstrated in Ref.[22], have not yet been systematically studied.

The polarization dynamics of the spin system of impurity-band electrons in longitudinal magnetic fields is affected, as follows from the above theory, by an exceedingly large number of factors. Electron-phonon, hyperfine, exchange and spin-orbit interactions are deeply involved. Magnetic field affects spin relaxation via splitting of spin levels and magnetic localization of electrons. As a result, some relaxation mechanisms are suppressed, and others (like relaxation in small clusters) may come into play. Under certain conditions, optical spin orientation in magnetic fields may bring about cooling of the electron spin system, and the decay of spin polarization will be governed by relaxation of energy rather than angular momentum. Disentangling this complex of phenomena requires systematic experiments in a range of impurity concentrations, with varying temperature, magnetic field and pump intensity, which has not yet been done. This is an obvious target for experimental research for the nearest future.

The quantitative understanding of the spin physics in impurity bands of bulk semiconductors, once it is reached, would be a good basis for studying spin systems of localized electrons in quantum wells and quantum-dot arrays. These nanostructures, attractive for the researchers from many points of view, unfortunately lack the uniformity of the localizing potential that greatly simplifies bringing together experiment and theory in bulk crystals. Possibly for this reason, interesting experimental results obtained in low-dimensional structures with localized interacting electrons remain so far disparate, and there is no general picture of spin dynamics in such systems. Hopefully, these difficulties will be soon overcome.

*Acknowledgements*

I am grateful to V.L.Korenev for very helpful discussions and comments. This work was supported by programs of Russian Academy of Sciences.